\documentclass[aps,twocolumn,showpacs]{revtex4}
\usepackage{amsmath}
\usepackage{amssymb}
\usepackage{graphicx}
\usepackage{dcolumn}

\begin{document}

\title{Resonance oscillations of magnetoresistance in double quantum wells}

\author{N. C. Mamani, G. M. Gusev, T. E. Lamas, A. K. Bakarov\footnote{Present address:
Institute of Semiconductor Physics, Novosibirsk 630090, Russia}}
\affiliation{Instituto de F\'{\i}sica da Universidade de S\~ao
Paulo, CP 66318 CEP 05315-970, S\~ao Paulo, SP, Brazil}
\author{O. E. Raichev}
\affiliation{Institute of Semiconductor Physics, National Academy of
Sciences of Ukraine, Prospekt Nauki 45, 03028, Kiev, Ukraine}
\date{\today}

\begin{abstract}
We present experimental and theoretical studies of the
magnetoresistance oscillations induced by resonance transitions
of electrons between tunnel-coupled states in double quantum wells.
The suppression of these oscillations with increasing temperature is
irrelevant to the thermal broadening of the Fermi distribution and
reflects the temperature dependence of the quantum lifetime of
electrons. The gate control of the period and amplitude of the
oscillations is demonstrated.
\end{abstract}

\pacs{73.21.Fg, 73.43.Qt}

\maketitle

\section{Introduction}

The Landau quantization$^1$ of electron states manifests itself
in oscillations of various physical quantities as functions of
the applied magnetic field. The magnetoresistance oscillations
in solids belong to the most fundamental quantum phenomena of this
kind$^2$. In particular, the Shubnikov-de Haas oscillations$^2$ (SdHO)
originating from sequential passage of the Landau levels through
the Fermi level, are always present in three-dimensional (3D)
and two-dimensional (2D) electron system with metallic type of
conductivity. Investigation of these oscillations is a powerful
source of information about band structure, Fermi surface, and
electron interaction mechanisms.

The SdHO are strongly damped with increasing temperature as the
thermal broadening of the Fermi distribution exceeds the cyclotron
energy. In quasi-2D electron systems, which are realized in quantum
wells with two (or more) occupied subbands, the possibility of
intersubband transitions of electrons leads to another kind
of magnetoresistance oscillations$^3$ which are not damped considerably
when the temperature raises. Such oscillations, known as the
magneto-intersubband oscillations (MISO), have been studied
in single quantum wells with two populated 2D subbands$^{3,4,5,6,7,8}$
and, theoretically, in 2D layers where electron states are split owing to
the spin-orbit interaction$^9$. The physical mechanism of the MISO is
the periodic modulation of the probability of intersubband transitions
by the magnetic field as the different Landau levels of the
two subbands sequentially come in alignment. Oscillations of similar
origin are present in 3D systems such as layered organic conductors
with weak coupling between the layers.$^{10,11,12}$

As compared to the SdHO, the MISO show a stronger suppression by the
disorder, so their observation requires clean (high-mobility) samples.
A decrease of the MISO amplitude with increasing temperature reflects
temperature dependence of the quantum lifetime of electrons. Experimental
studies of the MISO allow one to determine the quantum lifetimes
in the region of temperatures where the SdHO disappear completely.
Another advantage of the MISO is the possibility of direct
and precise determination of the intersubband energy gap, which
otherwise can be found from an analysis of the double periodicity of the
SdHO observed at low temperatures in various quasi-2D electron systems
such as single quantum wells with two populated 2D subbands,$^{4,6}$ double
quantum wells,$^{13,14}$ and 2D systems with spin-orbit splitting.$^{15,16}$

The double quantum wells (DQWs), which consist of two quantum wells
separated by a barrier, and where the electrons occupy two 2D subbands
coupled by tunneling, should be recognized as the most convenient system
for studying the MISO phenomenon, owing to the unique possibility to
control both intersubband energy gap and probabilities of intersubband
transitions in a wide range by using barriers of different
widths and biasing the structure by external gates. Nevertheless, no
studies of the MISO have been reported so far for these particular
systems. In this paper we report on the first observation and systematic
investigation of the MISO in symmetric high-mobility ($\mu \approx 10^{6}$
cm$^{2}$/Vs) GaAs DQWs with different barrier widths. Along with the
experimental results, we present a theoretical description of magnetotransport
in DQWs, which explains the results of our measurements.

The paper is organized as follows. The experimental results are described
in Sec. II. The theoretical description and the discussion of the results
are presented in Sec. III. The concluding remarks are given in the last section.

\section{Experiment}

\begin{table}
\label{tableI}
\caption{The sample parameters. $d_{W}$ is the well width, $d_{b}$ the
barrier thickness, $n_s$ the electron density, $\mu$ the zero
field mobility. $\Delta_{SAS}^{theor}$ is the symmetric-antisymmetric splitting
energy determined from self-consistent calculations, and $\Delta_{SAS}^{exp}$
is determined from the periodicity of the oscillations observed.}
\begin{tabular}{ccccccc}
\hline $d_{b}$ && $d_{W}$ & $n_{s}$ & $\mu$
&$\Delta_{SAS}^{theor}$&$\Delta_{SAS}^{exp}$ \\
\hline({\AA})&&({\AA})
& ($10^{11}$ cm$^{-2}$) & ($10^{3}$ cm$^{2}$/Vs) &(meV)&(meV) \\
\hline 14 & &140&9.32 &970 &3.87&4.22\\
20 &
&140&9.8 &900 &2.59&2.07\\
31& &140&9.19 &870&1.24&1.38 \\
50& &140&8.43 &830 &0.325&0.92\\
\hline
\end{tabular}
\end{table}

The samples are symmetrically doped GaAs DQWs with
equal widths $d_{W}=14$ nm separated by Al$_{x}$Ga$_{1-x}$As barriers of
different thickness $d_{b}$ varied from 1.4 to 5 nm. The samples have high
total sheet electron density $n_s \approx 9 \times10^{11}$ cm$^{-2}$
($4.5 \times 10^{11}$ cm$^{-2}$ per one layer). Both layers are shunted
by ohmic contacts. The densities in the wells are variable by a gold top
gate. The voltage of the gate, $V_g$, changes the density of the well
closest to the sample surface, with carrier density in the other well
being almost constant. The system is balanced (has equal densities in the
wells) at $V_g=0$.
The sample parameters are shown in Table 1. Over a dozen
specimens of both the Hall bar and van der Pauw geometries from four wafers
have been studied. We measure both longitudinal and Hall resistances at the
temperatures $T$ from 0.3 to 50 K and magnetic fields $B$, directed
perpendicular to the well plane, up to 12 T using conventional ac-locking
techniques with a bias current of 0.1-1 $\mu$A.

\begin{figure}[ht]
\includegraphics[width=8cm]{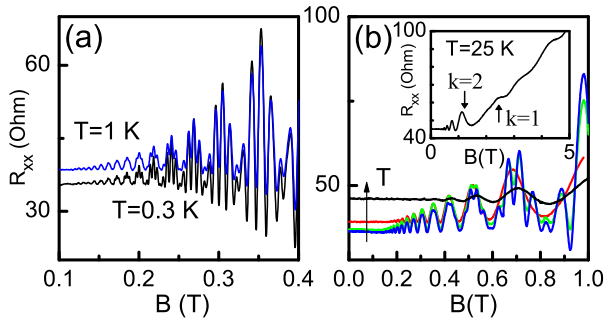}
\caption{(Color online) (a) Magnetoresistance oscillations in double
wells with barrier thickness $d_{b}=14$ {\AA} at $T=0.3$ K and $T=1$ K. The curves are
shifted for clarity. (b) Magnetoresistance oscillations at various temperatures
$T$ (K): 18, 10, 8, 4.2. The insert shows high-temperature magnetoresistance in
a wide magnetic field interval, the arrows indicate calculated positions of two
last peaks, $k=\Delta_T/\hbar \omega_{c}$.}
\end{figure}

Figure 1 shows the low-field part of the magnetoresistance
$R_{xx}(B)$ for the sample with $d_{b}=1.4$ nm and balanced
densities at different temperatures. At $T=0.3$ K we see
double-periodic SdHO in the form of beating pattern,$^{13,14}$ since
the energy splitting of 2D subbands in DQWs is small in comparison
to the Fermi energy. With the increase of $T$ the small-period SdHO
are strongly suppressed and the remaining long-period oscillations
can be identified as the MISO, by analyzing their $1/B$ periodicity
and temperature dependence as described below. First, assuming that
the magnetoresistance maxima correspond to the Landau-level alignment
condition$^5$ $\Delta_T = k \hbar \omega_{c}$, where $\Delta_T$ is the
intersubband splitting energy ($\Delta_T=\Delta_{SAS}$ for balanced
DQWs), $\omega_{c}=|e|B/mc$ is the cyclotron frequency ($e$ is the
electron charge, $m$ the effective mass, and $c$ the velocity of light),
and $k$ is an integer, we find $\Delta_T=4.22$ meV, which is close
enough to the calculated value of $\Delta_{SAS}$ (see Table 1).
Next, the amplitudes of the long-period oscillations, in contrast
to the SdHO amplitudes, are saturated at low temperatures. The dependence
of $R_{xx}(B)$ in the temperature interval $4.2 < T < 25$ K [Fig. 1 (b)]
shows that the oscillations are present at high temperatures. The amplitudes
of the oscillations are up to $30\%$ of the total resistance, so these
oscillations cannot be attributed to magnetophonon resonances whose
relative amplitudes are much smaller.$^{17}$

\begin{figure}[ht]
\includegraphics[width=6cm]{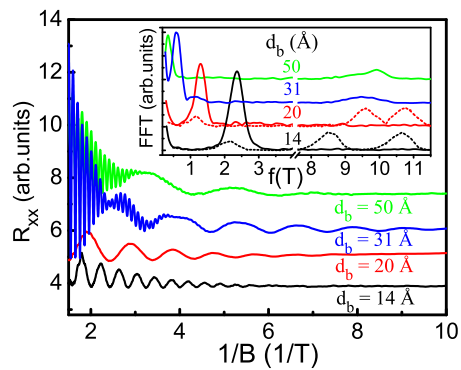}
\caption{(Color online)  Magnetoresistance of double wells with
various barrier widths versus inverse magnetic field, measured
at $T=4.2$ K for the samples with $d_{b}=14$ {\AA} and 20 {\AA} and
at $T=1.6$ K for the samples with $d_{b}=50$ {\AA} and 31 {\AA}. The
curves are shifted for clarity. The insert shows the Fourier power
spectra of $R_{xx}$, the additional dashed curves correspond to
measurements at $T=50$ mK and demonstrate two SdHO frequencies
for two subbands.}
\end{figure}

We have checked the $1/B$ periodicity of the oscillations for
the structures with different barriers. Figure 2 shows the
magnetoresistance traces versus $1/B$, while the results of the fast
Fourier transform (FFT) of the magnetoresistance data are shown in
the insert. The peak corresponding to the main period shifts
towards lower frequencies with increasing $d_{b}$, and the frequency
of the FFT peaks has nearly exponential dependence
on $d_{b}$. Since the splitting energy in the balanced
DQWs is known to decrease exponentially with increasing barrier
width, this observation is another confirmation of the
intersubband origin of the oscillations.

\begin{figure}[ht]
\includegraphics[width=6cm]{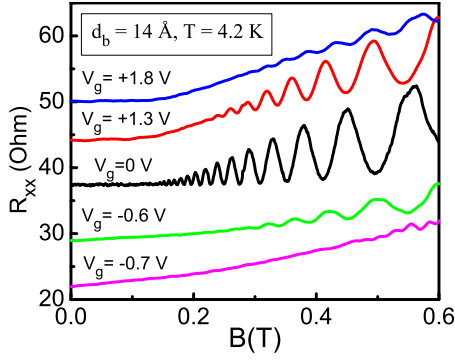}
\caption{(Color online) Low field magnetoresistance
oscillations for double wells with barrier width $d_{b}=14$ {\AA}
as a function of magnetic field for different gate voltages.
The curves are shifted for clarity. The corresponding areal
densities $n_s$ are 14, 12, 9.3, 8, and 7 $\times 10^{11}$
cm$^{-2}$ (from up to down).}
\end{figure}

Figure 3 shows the low-field part of the magnetoresistance for the sample
with $d_{b}=1.4$ nm at various fixed gate voltages away from the symmetry
point at $V_{g}=0$ V. As the absolute value of $V_g$ increases, the
oscillations are suppressed and their frequency becomes higher.
The positive magnetoresistance is clearly seen in these plots.

\section{Theory and discussion}

The mechanism responsible for the oscillations observed relies
on intersubband scattering, resonantly modulated by Landau
quantization. A quantitative description of magnetotransport in
DQWs taking into account elastic scattering of electrons is based on
the Hamiltonian written as the following $2 \times 2$ matrix in the
single-well eigenstate basis:$^{18}$
\begin{equation}
\hat{H}= \left| \begin{array}{cc} \mbox{\boldmath $\pi$}^2/2m + \Delta/2 + V^u_{\bf r}
& -\Delta_{SAS}/2 \\ -\Delta_{SAS}/2 &
\mbox{\boldmath $\pi$}^2/2m -\Delta/2 + V^l_{\bf r}\end{array} \right|,
\end{equation}
where $\mbox{\boldmath $\pi$}=-i\hbar {\bf \nabla}_{\bf r}-(e/c){\bf A}_{\bf r}$
is the kinematic momentum operator,
${\bf A}_{\bf r}$ is the vector potential describing the magnetic field, $V^i_{\bf r}$
are the random potentials of impurities and other inhomogeneities in the upper
($i=u$) and lower ($i=l$) well, and ${\bf r}=(x,y)$
is the in-plane coordinate. Next, $\Delta_{SAS}$ denotes the subband
splitting owing to the tunnel coupling, and $\Delta$ is
the splitting in the absence of tunneling. In symmetric (balanced) DQWs
$\Delta=0$. According to Eq. (1), the free-electron energy spectrum is given
by two sets of Landau levels of the upper ($+$) and lower ($-$)
subbands: $\varepsilon_n^{\pm}=\pm \Delta_T/2 + \varepsilon_n$, where
$\varepsilon_n= \hbar \omega_c (n+1/2)$ and $\Delta_{T}= \sqrt{\Delta_{SAS}^2+
\Delta^2}$ is the total intersubband gap. In symmetric DQWs $\Delta_{T}=
\Delta_{SAS}$ is the energy gap between the states
with symmetric and antisymmetric wave functions $\psi^{{\scriptscriptstyle (-)}}_z$
and $\psi^{{\scriptscriptstyle (+)}}_z$. Since we consider weak enough magnetic
fields, the Zeeman splitting is neglected.

Calculation of the diagonal ($\sigma_{d}$) and non-diagonal ($\sigma_{\bot}$)
components of conductivity tensor uses the Kubo formalism and Green's
function technique in the Landau-level representation.$^{19}$ In the model
of short-range symmetric scattering potential characterized by the correlator
$\langle \langle V^i_{\bf r} V^{i'}_{\bf r'} \rangle \rangle \simeq w
\delta_{ii'} \delta({\bf r}-{\bf r'})$, and in the limit of weak disorder, when
the scattering rate $1/\tau=mw/\hbar^3$ is small in comparison to $\Delta_T/\hbar$,
one has
\begin{equation}
\sigma_{d}=\frac{e^2 \omega_c}{\pi^2}
\int d\varepsilon \left( -\frac{\partial f_{\varepsilon}}{\partial
\varepsilon }\right) \left[\Phi^{{\scriptscriptstyle +}}_{\varepsilon}+
\Phi^{{\scriptscriptstyle -}}_{\varepsilon} \right],
\end{equation}
and $\sigma_{\bot}=|e|c n_s/B + \delta \sigma_{\bot}$ with
\begin{equation}
\delta \sigma_{\bot}= - \frac{2e^2}{\pi^2 \hbar}
\int d\varepsilon \left( -\frac{\partial f_{\varepsilon}}{\partial
\varepsilon }\right)
\left[ \Sigma^{{\scriptscriptstyle (+)}''}_{\varepsilon}
\Phi^{{\scriptscriptstyle +}}_{\varepsilon}
+ \Sigma^{{\scriptscriptstyle (-)}''}_{\varepsilon}
\Phi^{{\scriptscriptstyle -}}_{\varepsilon} \right],
\end{equation}
where $f_{\varepsilon}$ is the equilibrium Fermi distribution and
\begin{equation}
\Phi^{{\scriptscriptstyle \pm}}_{\varepsilon} \! = \! \frac{2 \Sigma^{{
\scriptscriptstyle (\pm)}''}_{\varepsilon}  \!
S^{{\scriptscriptstyle (\pm)}''}_{\varepsilon}  \! [\varepsilon \! - \!
\Sigma^{{\scriptscriptstyle (\pm)}'}_{\varepsilon}]}{
(\hbar \omega_c)^2+(2 \Sigma^{{\scriptscriptstyle (\pm)}''}_{
\varepsilon})^2},~S^{{\scriptscriptstyle (\pm)}}_{\varepsilon} \! \! = \!
\sum_{n=0}^{\infty} \frac{1}{\varepsilon \! - \! \varepsilon_n \! -
\! \Sigma^{{\scriptscriptstyle (\pm)}}_{\varepsilon} }.
\end{equation}
The quantities $\Sigma^{{\scriptscriptstyle (\pm)}}_{\varepsilon}$
are the self-energies for $+$ and $-$ states. Their real
($\Sigma^{{\scriptscriptstyle (\pm)}'}_{\varepsilon}$) and imaginary
($\Sigma^{{\scriptscriptstyle (\pm)}''}_{\varepsilon}$) parts can be
found using the self-consistent Born approximation$^{20}$ generalized
for DQWs:
\begin{equation}
\Sigma^{{\scriptscriptstyle (\pm)}}_{\varepsilon}=\pm \frac{\Delta_T}{2} + \!
\frac{\hbar^2 \omega_c}{4 \pi \tau} \left[\left(1 \! \pm \! \delta^2 \right)
S^{{\scriptscriptstyle (+)}}_{\varepsilon} + \! \left(1 \! \mp \! \delta^2 \right)
S^{{\scriptscriptstyle (-)}}_{\varepsilon} \right],
\end{equation}
where $\delta^2=(\Delta/\Delta_T)^2$. The energies $\pm \Delta_T/2$
are included in the self-energies for notational convenience.
Equations (2)-(5) give the full description of magnetotransport in DQWs
with symmetric short-range scattering within the SCBA.

In weak magnetic fields, both $\Sigma^{{\scriptscriptstyle (\pm)}}_{\varepsilon}$ and
$S^{{\scriptscriptstyle (\pm)}}_{\varepsilon}$ can be represented as serial expansions
in small Dingle factors $e^{-\alpha}$, where $\alpha=\pi/\omega_c \tau$, see Ref. 20.
This procedure gives $$S^{{\scriptscriptstyle (\pm)}}_{\varepsilon}=\frac{i \pi}{\hbar \omega_c}
\left\{ 1 - 2 e^{-\alpha} \exp \left[-i \frac{2 \pi (\varepsilon \mp \Delta_T/2)}{\hbar
\omega_c} \right] + \ldots \right\}$$ and leads to the following expression for the resistivity
$\rho_{xx}=\sigma_{d}/(\sigma_{d}^2+\sigma_{\bot}^2)$:
\begin{eqnarray}
\rho_{xx} \simeq \rho_0 \left\{ 1-4 e^{-\alpha} {\cal T}
\cos \left( \frac{2 \pi \varepsilon_F}{\hbar \omega_c} \right)
\cos \left( \frac{\pi \Delta_T}{\hbar \omega_c} \right) \right. \nonumber \\
\left.
+ e^{-2 \alpha} \left[a_+ + a_-
\cos\left( \frac{2 \pi \Delta_T}{\hbar \omega_c} \right)\right] \right\},
\end{eqnarray}
where $\rho_0=m/e^2 n_s \tau$ is the zero-field Drude resistivity, and
$a_{\pm}=1 \pm \delta^2 -(1 \pm \delta^4)/[1+(\omega_c\tau)^2]$.
The Fermi energy $\varepsilon_F$ is counted from the middle point
between the subbands, and the function ${\cal T}=(2 \pi^2 T/\hbar
\omega_c)/\sinh(2 \pi^2 T/\hbar \omega_c)$ describes thermal
suppression of the resistivity oscillations. The second
($\propto e^{-\alpha}$) term in Eq. (6) describes usual SdHO
with beatings, while the last ($\propto e^{-2 \alpha}$) term
contains the MISO contribution which does not have the thermal
suppression (the temperature-dependent corrections to this
term are neglected because they are small in comparison to the
second term). The expression for the coefficients $a_{\pm}$
can be rewritten in terms of intersubband and intrasubband
scattering times in DQWs, which are given by $\tau_{inter}=2\tau/(1-
\delta^2)$ and $\tau_{intra}=2 \tau/(1+\delta^2)$, respectively,
so that $\tau^{-1}=\tau^{-1}_{intra}+ \tau^{-1}_{inter}$.
This representation shows that the MISO contribution (at the
coefficient $a_{-}$) is proportional to the intersubband
scattering rate $\tau^{-1}_{inter}$.

As the temperature raises, the second term in the right-hand side
of Eq. (6) vanishes, and only the MISO remain. These oscillations are
periodic in $1/B$ and have maxima under the condition
$\Delta_T = k \hbar \omega_{c}$. In balanced DQWs ($\delta=0$ and
$a_+=a_-$), when the electron density distributions
$|\psi^{{\scriptscriptstyle (-)}}_z|^2$ and $|\psi^{{\scriptscriptstyle (+)}}_z|^2$
for $-$ and $+$ states are equal, the probability of intersubband scattering
is high. As the system is driven out of the balance by the gates
($\delta^2$ increases), the wave functions for $-$ and $+$ subbands
are localized in the different wells, and
intersubband scattering is suppressed, so the MISO amplitude
(coefficient $a_-$) is reduced, while the non-oscillating
positive manetoresistance (coefficient $a_+$) becomes larger.
This behavior is seen in experiment (Fig. 3), along
with enhanced frequency of the oscillations owing to larger
intersubband separation. Equation (6) shows that the resistivity
in the minima, when $\Delta_T = (k+1/2) \hbar
\omega_c$, is equal to $\rho_0$ for balanced DQWs in low magnetic
fields. This feature, also seen in our experiment, is described
by the fact that the total density of states,
$(m \omega_c/\pi^2 \hbar) [S^{{\scriptscriptstyle (+)}''}_{\varepsilon}
+ S^{{\scriptscriptstyle (-)}''}_{\varepsilon}]$, given by a
superposition of the densities of states in $+$ and $-$ subbands,
does not oscillate with energy under the condition $\Delta_T =(k+1/2)
\hbar \omega_c$, so the magnetic field has no effect on the
scattering and resistivity under this condition.

\begin{figure}[ht]
\includegraphics[width=6cm]{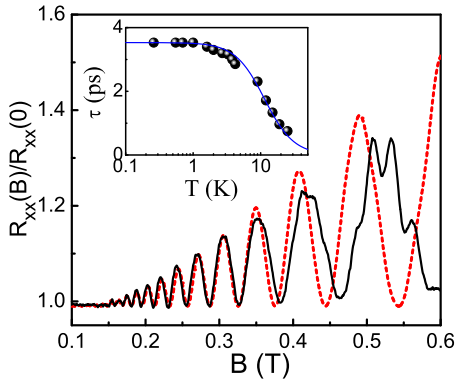}
\caption{(Color online)  Comparison of the experimental (solid line)
and theoretical (dashed line) magnetoresistance for balanced DQWs
with $d_{b}=14$ {\AA} at 4.2 K. The theoretical curve is slightly
shifted down to account for a weak negative magnetoresistance at
low fields. The insert shows the temperature dependence of the quantum
lifetime of electrons extracted from the amplitude fit of the MISO
(circles), and the theoretical dependence assuming $T^2$ scaling
of $1/\tau$ (line).}
\end{figure}

For a more detailed comparison of theory and experiment, we have
carried out a fit of the magnetoresistance (6) to the data obtained
for the balanced DQWs with $d_{b}=1.4$ nm at $T=4.2$ K. In order to
fit the MISO frequency at low $B$, we have taken $\Delta_T=4.22$
meV, while the amplitude fit has been done assuming that the ratio
of the transport time $\tau_{tr}=37$ ps determined from the mobility
to the quantum lifetime $\tau$ standing in the Dingle factor is
equal to 12 (large ratios of $\tau_{tr}$ to $\tau$
are normally met in modulation-doped heterostructures). With these
parameters used, we obtain an excellent agreement of theory and
experiment in the region of magnetic fields up to 0.4 T; see Fig. 4.
A good agreement is also obtained for lower temperatures, when the
SdHO are present. Above 0.4 T, the observed MISO frequency becomes
slightly smaller than expected from the theory. We do not know exact
origin of this discrepancy, and assume that it is possibly related
to modification of the tunnel coupling by the magnetic field.$^{21}$

The amplitude fit of the MISO in a wide region of temperatures
allows us to plot the temperature dependence of the quantum lifetime
of electrons. We have found that $\tau$ is constant at $T < 1$ K and
decreases at higher temperatures. This decrease cannot be attributed to
quasielastic scattering of electrons by acoustic phonons, because the
acoustic-phonon scattering rates in GaAs quantum wells are more than
one order smaller in the described temperature region. On the other hand,
the consideration of electron-electron scattering in DQWs$^{18,22}$ gives
a reasonable explanation of the observed temperature behavior. To prove
this, we have compared the experimental dependence shown in the insert
to Fig. 4 with the theoretically predicted dependence$^{18,22}$
$\tau(T)=\tau(0)/[1+ \lambda (T/T_0)^2 ]$, where
$T_0=\sqrt{\varepsilon_F \hbar/\tau(0)}$ and $\lambda$
is a numerical coefficient of the order of unity, considered here as a single
fitting parameter. Using the experimentally determined values $\varepsilon_F
\simeq 17$ meV and $\hbar/\tau(0) \simeq 0.2$ meV, we have found a good
agreement between experiment and theory in the whole temperature region
for $\lambda=3.5$.

\section{Conclusions}

In conclusion, we have observed and investigated the magnetoresistance
oscillations caused by resonant modulation of elastic intersubband scattering
in DQWs by Landau quantization. These oscillations survive at high temperatures
and represent a prominent and well-reproducible feature of magnetotransport
in high-mobility DQWs.

The oscillations we observe are caused by the same physical mechanism
as the magneto-intersubband oscillations (MISO) previously studied in
single quantum wells with two populated subbands. However, there
are qualitatively different features in manifestation of the MISO
phenomenon in DQWs. First, since the Fermi energy in DQWs is typically
much larger than the subband separation, the MISO have a large period
as compared to the SdHO period, and they are clearly identifiable without
an additional Fourier analysis. For the same reason, one can observe
all intersubband resonances up to the last one, when the cyclotron energy
equals the subband separation, in relatively weak magnetic fields
(see Fig. 1 b). Next, since the intesubband scattering is DQWs is easily
controllable by modifying the wave functions of electrons via the gate,
the MISO appear to be very sensitive to the gate voltage. We also
point out that observation of the large-amplitude MISO requires large
quantum lifetimes and high probability of intersubband scattering.
These two requirements are difficult to satisfy simultaneously
in single quantum wells with two populated subbands, because large
quantum lifetimes are usually attainable in the modulation-doped
systems, where the scattering probability strongly decreases with
increasing momentum transfer, and the intersubband scattering, which
requires a large momentum transfer, is suppressed.$^5$ In contrast,
the intersubband scattering in DQWs does not require a large momentum
transfer, and its probability is comparable to the probability of
intrasubband scattering.

Theoretical consideration of the magnetotransport in DQWs leads to an
analytical expression for the oscillating resistivity which survives at
high temperatures, the last term in Eq. (6). This expression, as far as we
know, has not been derived previously. Its simple form is a consequence
of symmetric scattering, when the subbands are characterized by a single
quantum lifetime $\tau$, common for both subbands. The quantum lifetime
can be extracted from the SdHO amplitudes at low temperatures, while the
measurements of the MISO amplitudes allow us to find its temperature
dependence in a wide region of temperatures. This dependence becomes an
independent confirmation of the influence of electron-electron scattering
on the quantum lifetime of 2D electrons.

We acknowledge fruitful discussions with S. Sokolov, N. Studart, and
O. Balev. This work was supported by FAPESP and CNPq (Brazilian
agencies).

\end{document}